\documentclass[12pt,a4paper]{article}
\usepackage{amssymb}
\usepackage{amsmath}
\RequirePackage{graphicx}
\RequirePackage{mathptmx}      
\RequirePackage{flushend}
\RequirePackage[numbers,sort&compress]{natbib}
\usepackage[colorlinks]{hyperref}
\hypersetup{
     breaklinks=true,
     pdfstartview={FitH},    
    colorlinks=true,       
    linkcolor=blue,          
    citecolor=red,        
    filecolor=magenta,      
    urlcolor=Green,           
    anchorcolor=blue,      
    linktocpage=true
}
\usepackage{psfig}
\begin{document}
\pagestyle{myheadings}
\pagenumbering{arabic}
\newcommand{\be}{\begin{equation}}
\newcommand{\ee}{\end{equation}}
\newcommand{\bea}{\begin{eqnarray}}
\newcommand{\eea}{\end{eqnarray}}
\newcommand{\bc}{\begin{center}}
\newcommand{\ec}{\end{center}}
\begin{center}{\Large\bf Study of parametrized dark energy models with a general non-canonical scalar field}
\\[15mm]
Abdulla Al Mamon\footnote{E-mail : abdullaalmamon.rs@visva-bharati.ac.in}~and
Sudipta Das\footnote{E-mail:  sudipta.das@visva-bharati.ac.in}\\
{\em Department of Physics, Visva-Bharati,\\
Santiniketan- 731235, ~India.}\\
[15mm]
\end{center}
\vspace{0.5cm}
\vspace{0.5cm}
\begin{abstract}
In this paper, we have considered various dark energy models in the framework of a non-canonical scalar field with a Lagrangian density of the form\\
${\cal L}(\phi , X)=f(\phi)X{\left(\frac{X}{M^{4}_{Pl}}\right)}^{\alpha -1} - V(\phi)$, which provides the standard canonical scalar field model for $\alpha=1$ and $f(\phi)=1$. In this particular non-canonical scalar field model, we have carried out the analysis for $\alpha=2$. We have then obtained cosmological solutions for constant as well as variable equation of state parameter ($\omega_{\phi}(z)$) for dark energy. We have also performed the data analysis for three different functional forms of $\omega_{\phi}(z)$ by using the combination of SN Ia, BAO and CMB datasets. We have found that for all the choices of $\omega_{\phi}(z)$, the SN Ia $+$ CMB/BAO dataset favors the past decelerated and recent accelerated expansion phase of the universe. Furthermore, using the combined dataset, we have observed that the reconstructed results of $\omega_{\phi}(z)$ and $q(z)$ are almost choice independent and the resulting cosmological scenarios  are in good agreement with the $\Lambda$CDM model (within the $1\sigma$ confidence contour). We have also derived the form of the potentials for each model and the resulting potentials are found to be a quartic potential for constant $\omega_{\phi}$ and a polynomial in $\phi$ for variable $\omega_{\phi}$.
\end{abstract} 
\section{Introduction}
One of the biggest challenges in modern cosmology is understanding the nature of the dark energy (DE), which seems to be responsible for the observed accelerated expansion phase of the universe at the present epoch \cite{acc1gnc,acc2gnc}. Among the many candidates for DE, the cosmological constant ($\Lambda$) emerges as the most natural and the simplest possibility. However, $\Lambda$-cosmology suffers from the so-called ``{\it fine tuning}" and ``{\it cosmic coincidence"} problems \cite{gncccp1,gncccp2}. These theoretical problems motivated cosmologists to think beyond the cosmological constant and explore other unknown components which may be responsible for the late-time accelerated expansion phase of the universe. The scalar field models have played an important leading role as a candidate of DE due to its dynamical nature and simplicity. Till now, a variety of scalar field DE models have been proposed, such as quintessence (canonical scalar field), k-essence, phantom, tachyon, dilatonic dark energy and so on (for details, see review \cite{revnc0} and the references therein). But, the origin and nature of DE still remains completely unknown despite many years of research. \\
\par It is strongly believed that the universe had a rapid exponential expansion phase during a short era in the very early epoch. This is known as {\it inflation} \cite{aginfgnc,gncpotm1} which can give satisfactory explanation to the problems of the Hot Big Bang cosmology (for example, horizon, flatness and monopole problems). Generally, cosmologists realized this inflationary scenario by using a single canonical scalar field called ``{\it inflaton}", which has a canonical kinetic energy term ($\frac{{\dot{\phi}}^{2}}{2}$) in the Lagrangian density. In the literature, there also exists some inflationary models in which the kinetic energy term is different from the standard canonical scalar field case (instead of the standard form $\frac{{\dot{\phi}}^{2}}{2}$). Such models are commonly known as the {\it non-canonical scalar field models of inflation}. Such non-canonical scalar fields have been found to have many attractive features compared to the canonical scalar field case, for example, the slow-roll conditions can be achieved more easily as compare to the canonical case.  Many interesting possibilities with these models have been recently studied in the literature (see refs \cite{nc1gnc,nc2gnc,nc3gnc,unnignc,nc4gnc,nc5gnc,nc6gnc,nc7gnc,nc8gnc,
nc9gnc,nc10gnc,nc11gnc,nc12gnc,fanggnc,sdgnc,aamgnc}). It has been first shown in refs. \cite{nc1gnc,nc2gnc} that k-essence model (which is an important class of non-canonical scalar field model) is capable of generating inflation in the early epoch. Later, Chiba et al. \cite{nc3gnc} showed that such models can equally effectively describe a DE scenario. Since the nature of DE is completely unknown, it is quite reasonable to consider a non-canonical scalar field as a candidate for DE component and check for the viability of such models. Within the framework of a non-canonical scalar field, in this work, we shall try to obtain an observationally viable cosmological model to analyze the behavior of the deceleration parameter ($q$) and the equation of state (EoS) parameter ($\omega_{\phi}$) for describing the expansion history of the universe. The motivation for this work has been discussed in detail in section 2.  As already mentioned, as the nature of DE is unknown to us, we eventually have no firm idea regarding whether the EoS parameter of DE is a constant quantity or whether it is dynamical in nature. In this connection, the most effective choice is to assume a specific functional form for the dark energy EoS parameter $\omega_{\phi}$ as a function of the redshift $z$ (for detail see section \ref{gncsecm2}). To study the non-canonical scalar field DE model in a more general framework, in this paper, we have considered both the possibilities. First, we have studied the model for a constant EoS parameter $\omega_{\phi}$, which is in the range $-1<\omega_{\phi}<-\frac{1}{3}$ so as to obtain acceleration. Secondly, we have considered three different choices for $\omega_{\phi}(z)$ in order to cover a wide range of the DE evolution. We have then solved the field equations and analyzed the respective cosmological scenarios for all the cases. For all the models, the deceleration parameter $q$ is found to exhibit an evolution from early deceleration to late time acceleration phase of the universe. This feature is essential for the structure formation of the universe. For all the models, we have also derived the potential $V(\phi)$ in terms of the scalar field $\phi$ by considering a specific parametrization of $f(\phi)$. In order to compare the theoretical models of DE (for dynamical $\omega_{\phi}$) with the observations, we have used the SN Ia, BAO and CMB dataset to constrain the various model parameters (for details see Appendix A). We have found that the combined dataset favors the $\Lambda$CDM model within the $1\sigma$ confidence contour. We have given the detail results of this work in section \ref{gncresults}.\\
\par The present paper is organized in the following way. In the next section, we have introduced some basic equations of a general non-canonical scalar field model and also discussed the motivation of this work. We have then obtained the general solutions of the field equations for a particular choice of the function $f(\phi)$ and for different forms of the EoS parameter $\omega_{\phi}$. In section \ref{gncresults}, we have summarized the results of this work. Finally, the conclusions of this work are presented in section \ref{gncconclusions}. Additionally, for completeness, we have performed the combine data analysis in Appendix A and found the observational constraints on $\omega_{\phi}(z)$ and $q(z)$ using the SN Ia, BAO and CMB datasets. 
\section{Basic Framework}
Usually, the scalar field models are characterized by a general action which has the following functional form
\be\label{action}
S={\int d^{4}x \sqrt{-g}{\left(\frac{R}{2} + {\cal L}(\phi,X)\right)}} + S_{m}
\ee
where $R$ is Ricci scalar, ${\cal L}(\phi,X)$ is the Lagrangian density which is an arbitrary function of the scalar field $\phi$ and its kinetic term $X$. The kinetic term $X$ is defined as, $X=\frac{1}{2}\partial_\mu \phi \partial^\mu \phi$, which is a function of time only. The last term $S_{m}$ represents the action of the background matter. Throughout this paper we shall work in natural units, such that $8\pi G=c=1$.\\ 
The expressions for the energy density ($\rho_{\phi}$) and pressure ($p_{\phi}$) associated with the scalar field are given by
\bea
\rho_{\phi}=2X\frac{\partial {\cal L}}{\partial X}- {\cal L}\\
p_{\phi}={\cal L}(\phi,X)
\eea
where $X=\frac{1}{2}{\dot{\phi}}^{2}$. 
\par In general, the Lagrangian for a scalar field model can be
represented as (Melchiorri et al. \cite{genncgnc})
\be\label{geneq}
{\cal L}(\phi , X)=f(\phi)F(X)-V(\phi)
\ee
where $f(\phi)$ and $F(X)$ are arbitrary functions of $\phi$ and $X$ respectively. $V(\phi)$ is the potential for the scalar field $\phi$.\\
\par Let us consider a homogeneous, isotropic and spatially flat
FRW universe which is characterized by the following line element
\be\label{frw}
ds^{2}=dt^{2}-a^{2}(t){\left[dr^{2} + r^{2}d{\theta}^{2} + r^{2}sin^{2}\theta d{\phi}^{2}\right]}
\ee
where $a(t)$ is the scale factor of the universe.\\
With the FRW geometry, the equations of motion take the form
\be\label{eq1}
  3H^{2}= 2f(\phi)X F_{X} -f(\phi)F + V(\phi) + \rho_{m} 
\ee
\be\label{eq2}
 \dot{H}= - \frac{1}{2} {\left[ 2f(\phi)XF_{X}  +  \rho_{m} \right]}
\ee
\be\begin{split} &\label{eq3}
{\left[ f(\phi)F_{X} + 2f(\phi)XF_{XX} \right]}{\ddot{\phi}} +3Hf(\phi){\dot{\phi}}F_{X} + 2X\frac{\partial f(\phi)}{\partial \phi}F_{X}\\ & 
~~~~~~~~~~~~~~~~~~~~~~~~~~~~-\frac{\partial f(\phi)}{\partial \phi}F + \frac{\partial V(\phi)}{\partial \phi} = 0 
\end{split}
\ee
\be\label{eq4}
{\dot{\rho}}_{m} + 3H\rho_{m}=0
\ee
where $H=\frac{\dot{a}}{a}$ denotes the Hubble parameter, an overdot indicates differentiation with respect to the time-coordinate $t$ and $\rho_{m}$ represents the energy density of the matter component of the universe, $F_{X} \equiv \frac{\partial F}{\partial X}$ and $F_{XX}\equiv\frac{\partial^{2}F}{\partial X^{2}}$.\\
\par It deserves mention that equation (\ref{geneq}) includes all the popular single scalar field models. It reduces to canonical scalar field model when $f(\phi)={\rm constant}=1$ and $F(X)=X$. Again, it describes a pure k-essence model when $V(\phi)=0$ and a phantom scalar field model when $f(\phi)=1$ and $F(X)=-X$. It is interesting to note that equation (\ref{geneq}) reduces to {\it general non-canonical scalar field model} [${\cal L}(\phi , X)=F(X)-V(\phi)$] when $f(\phi)=1$. This type of non-canonical scalar field models were proposed by Fang et al. \cite{fanggnc}. They studied several aspects of this type of scalar fields for different forms of $F(X)$. Recently, these type of non-canonical scalar field models have gathered attention due to their simplicity. Unnikrishnan et al. \cite{unnignc} have showed that for non-canonical scalar field models, the slow-roll conditions can be more easily satisfied compared to the canonical inflationary theory. They have shown that such models (with quadratic and quartic potentials) are more consistent with the current observational constraints relative to the canonical inflation. They have also shown that such non-canonical models can drop the tensor-to-scalar ratio than their canonical counterparts. In fact, a lot of work have been done in the framework of non-canonical inflationary scenario in the early epoch \cite{nc1gnc,nc2gnc,nc3gnc,unnignc, nc4gnc,nc5gnc,nc6gnc,nc7gnc,nc8gnc,nc9gnc,nc10gnc,
nc11gnc,nc12gnc,fanggnc}. Furthermore, Franche et al. \cite{conditiongnc} showed that the non-canonical scalar fields are the most universal case with a general Lagrangian density satisfing certain conditions. These interesting properties of non-canonical scalar field models motivated us to study the cosmological aspects of such fields in a more general framework in the context of dark energy. In the literature, a large number of functional forms of ${\cal L}(\phi , X)$ have been proposed so far, see for example \cite{unnignc,alimgnc,mvgnc,unni1gnc}. In our earlier works \cite{sdgnc,aamgnc}, we have considered a Lagrangian density of the following form
\be\label{early}
{\cal L}(\phi , X)=X^{2} - V(\phi)
\ee 
which can be obtained from the general form of Lagrangian density \cite{unnignc,mvgnc,unni1gnc}
\be
{\cal L}(\phi , X)=X{\left(\frac{X}{M^{4}_{Pl}}\right)}^{\alpha -1} - V(\phi)
\ee
for $\alpha=2$ and $M_{Pl}=\frac{1}{\sqrt{8\pi G}}=1$. The above equation describes a purely canonical scalar field Lagrangian density [${\cal L}(\phi , X)=X - V(\phi)$] when $\alpha = 1$.\\
\par In this present work, we try to extend our previous works in \cite{sdgnc,aamgnc} by considering a general non-canonical scalar field model which has the following Lagrangian density
\be\label{eqnclagd}
{\cal L}(\phi , X)=f(\phi)X{\left(\frac{X}{M^{4}_{Pl}}\right)}^{\alpha -1} - V(\phi)
\ee
Here, following \cite{sdgnc,aamgnc} also, we consider $\alpha=2$ and $M_{Pl}=\frac{1}{\sqrt{8\pi G}}=1$.\\
In this case, the energy density and pressure of the scalar field are given by
\be\label{eqncrphi}
\rho_{\phi}=\frac{3}{4}f(\phi){\dot{\phi}}^{4} + V(\phi)
\ee
\be\label{eqncpphi}
p_{\phi}=\frac{1}{4}f(\phi){\dot{\phi}}^{4} - V(\phi)
\ee
It is evident from equations (\ref{eqncrphi}) and (\ref{eqncpphi}) that the usual definition of $\rho_{\phi}$ $(=\frac{1}{2}{\dot{\phi}}^2+V(\phi))$ and $p_{\phi}$ $(=\frac{1}{2}{\dot{\phi}}^2-V(\phi))$ for a standard canonical scalar field model gets modified due to the Lagrangian density (\ref{eqnclagd}). 
Also, the equations of motion for this Lagrangian (equations (\ref{eq1})-(\ref{eq4})) come out as
\be\label{eqf1}
  3H^{2}= \frac{3}{4}f(\phi){\dot{\phi}}^{4} + V(\phi) + \rho_{m} 
\ee
\be\label{eqf2}
 \dot{H}= - \frac{1}{2} {\left[ f(\phi){\dot{\phi}}^{4}  +  \rho_{m} \right]}
\ee
\be\label{eqf3}
{\dot{\rho}}_{\phi} + 3H(\rho_{\phi} + p_{\phi})=0
\ee
\be\label{eqf4}
{\dot{\rho}}_{m} + 3H\rho_{m}=0
\ee
The solution for $\rho_{m}$ from equation (\ref{eqf4}) is obtained as
\be
\rho_{m}(z)=\rho_{m0}(1+z)^{3}
\ee
where $\rho_{m0}$ is the matter density at the present time, $z=\frac{a_{0}}{a}-1$ is the redshift and the present value of the scale factor $a_{0}$ is normalized to unity.\\
One of the important quantities in cosmology is the dark energy EoS parameter $\omega_{\phi}=\frac{p_{\phi}}{\rho_{\phi}}$ which, in our case, is given by 
\be
\omega_{\phi}=\frac{f(\phi){\dot{\phi}}^{4} - 4V(\phi)}{3f(\phi){\dot{\phi}}^{4} + 4V(\phi)}
\ee
From equation (\ref{eqf3}),one can then obtain the expression for the energy density of the scalar field as
\be\label{eqncrhophigen}
\rho_{\phi}(z) =\rho_{\phi 0} {\rm exp}{\left[3\int^z_{0}{\frac{1+\omega_{\phi}(z^{\prime})}{1+z^{\prime}}dz^{\prime}}\right]}
\ee
where $\rho_{\phi 0}$ is an integration constant.\\
Now, the Friedmann equation can be written in the following integrated form
\be\label{eqnchub}
H^2(z)=H^2_{0}{\left[\Omega_{m0}(1+z)^3 + \Omega_{\phi 0}{\rm exp}{\left(3\int^z_{0}{\frac{1+\omega_{\phi}(z^{\prime})}{1+z^{\prime}}dz^{\prime}}\right)}\right]}
\ee
where $H_{0}$ is the Hubble parameter at the present epoch, $\Omega_{m0}=\frac{\rho_{m0}}{3H^{2}_{0}}$ and $\Omega_{\phi 0}(=\frac{\rho_{\phi 0}}{3H^{2}_{0}})=1-\Omega_{m0}$ are the density parameters of the matter and scalar field (or dark energy) respectively at the present epoch.\\ 
For this model, from equations (\ref{eqncrphi}) and (\ref{eqncpphi}), the potential can be expressed (as a function of redshift $z$) as
\be\label{eqncvzgen}
V(z)=\frac{1}{4} (1-3\omega_{\phi}(z)) \rho_{\phi}(z)
\ee
and
\be\label{eqncfphid4}
f(\phi){\dot{\phi}}^4 = (1+\omega_{\phi})\rho_{\phi}
\ee
\par In order to solve the field equations analytically we will proceed as follows. Out of equations (\ref{eqf1}), (\ref{eqf2}), (\ref{eqf3}) and (\ref{eqf4}), only three are independent equations in view of the Bianchi identity (with five unknown quantities $H$, $\rho_{m}$, $f(\phi)$, $V(\phi)$ and $\phi$). So naturally one has to assume two relationships among the different variables to solve the system of equations.\\
Following above argument, in this paper, we have assumed that the quantity $f$ has a functional form
\be\label{eqncfans}
f={\left({\frac{f_{0}}{H}}\right)}^4
\ee
where $f_{0}$ is an arbitrary constant. It deserves mention here that the above parametrization of $f(\phi)$ helps us to close the system of equations. With this input, the equation (\ref{eqncfphid4}) can be written in the following integral form
\be\label{eqncphiz}
\phi(z) = \phi_{0} + \int^z_{0} {F(z^{\prime})dz^{\prime}}
\ee
where $\phi_{0}$ is an arbitrary constant of integration and $F(z^{\prime}) = \frac{[(1+\omega_{\phi}(z^{\prime}))\rho_{\phi}(z^{\prime})]^{\frac{1}{4}}}{f_{0}(1+z^{\prime})}$.\\
Another important observable quantity, the deceleration parameter $q(z)$, can also be expressed in terms of $H(z)$ as 
\be\label{eqncqz}
q(z)= -\frac{\ddot{a}}{aH^2}=-1 + (1+z) \frac{d\ln{H(z)}}{dz}
\ee
which describes the evolution of our universe.\\
\par We shall now concentrate on the dark energy EoS parameter $\omega_{\phi}(z)$. If a function of $\omega_{\phi}(z)$ is given, then we can find the evolution of $\rho_{\phi}(z)$ from equation (\ref{eqncrhophigen}). As a result, we can also find the evolutions of $H(z)$, $q(z)$, $V(z)$ and $\phi(z)$. Inverting $\phi(z)$ into $z(\phi)$ and using equation (\ref{eqncvzgen}), one can then obtain the potential $V(\phi)$ in terms of $\phi$. As already mentioned, we have considered a specific parametrization of $f(\phi)$ and still we need another assumption to match the number of unknown parameters with the number of independent equations. With this freedom, we choose different functional forms for $\omega_{\phi}(z)$, the equation of state parameter. In the next section, we have tried to obtain the functional forms of various cosmological parameters for different choices of $\omega_{\phi}(z)$ and have studied their cosmological implications.
\section{Theoretical Models}\label{gnctheom}
In this section, we shall consider two phenomenological DE models for obtaining the current acceleration of the universe in the framework of a general non-canonical scalar field theory.
\subsection{Model-I: Accelerating universe driven by a constant EoS parameter for dark energy $(-1< \omega_{\phi} < -\frac{1}{3})$}
In this model, we shall investigate the properties of an accelerated expanding universe driven by a non-canonical scalar field dark energy with a constant EoS parameter. Recent observations suggest that the dark energy EoS $\omega_{\phi}$ is very close to $-1$ and the approximate bound on $\omega_{\phi}$ is $-1.1\le \omega_{\phi}\le -0.9$ \cite{wvgnc,tmdgnc}. Keeping this limit in mind, we choose a constant $\omega_{\phi}$ in the limit $-1< \omega_{\phi} < -\frac{1}{3}$, which ensures that the model does not deviate much from a $\Lambda$CDM model. \\
\par
\begin{figure}
\centering
\includegraphics[width=0.38\textwidth,height=0.22\textheight]{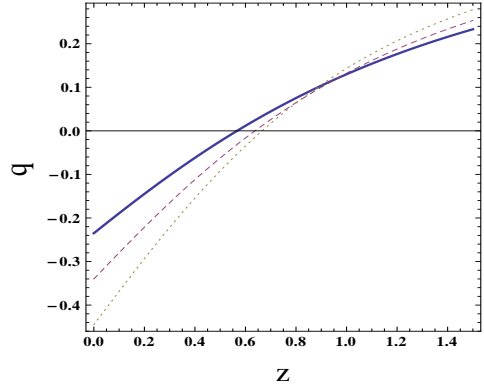}
\caption{\normalsize{\em Plot of $q$ vs. $z$ for $\omega_{\phi}=-0.7$ (thick curve), $\omega_{\phi}=-0.8$ (dashed curve) and $\omega_{\phi}=-0.9$ (dotted curve). For all these plots, we have taken $\Omega_{m0}=0.3$. The horizontal line is for $q(z)=0$.}}
\label{figncqm1}
\end{figure}

 In this case, the energy density of DE can be obtained as (from equation (\ref{eqncrhophigen}))
\be\label{eqrhozm0}
\rho_{\phi}(z)=\rho_{\phi 0}(1+z)^{3(1+\omega_{\phi})}
\ee
where $\rho_{\phi 0}$ is the integration constant which represents the dark energy density at the present time. Then the corresponding Friedmann equation becomes
\be
H^2(z)=H^{2}_{0}{\left[\Omega_{m0}(1+z)^{3} + (1-\Omega_{m 0})(1+z)^{3(1+\omega_{\phi})}\right]}
\ee
Inserting the Hubble parameter $H(z)$ (as given in the above equation) into equation (\ref{eqncqz}), we have obtained the deceleration parameter as
\be
q(z)=-1 + \frac{3 + 3\kappa_{1}(1+z)^{3\omega_{\phi}}}{2 + 2\kappa_{2}(1+z)^{3\omega_{\phi}}}
\ee
where $\kappa_{1}=\frac{(1-\Omega_{m 0})(1+\omega_{\phi})}{\Omega_{m 0}}$ and $\kappa_{2}=\frac{1-\Omega_{m 0}}{\Omega_{m 0}}$. The plot of $q(z)$ against $z$ is shown in figure \ref{figncqm1} for different values of $\omega_{\phi}$ (within the range,  $-1<\omega_{\phi}<-\frac{1}{3}$) and $\Omega_{m 0}=0.3$. Figure \ref{figncqm1} shows that $q(z)$ crosses its transition point from its positive value regime to negative value regime in the recent past, which is consistent with the independent measurements reported by several authors (see ref. \cite{aamgnc2} and the references therein).\\ 
With the help of equations (\ref{eqncphiz}) and (\ref{eqrhozm0}), one can solve for the scalar field in the flat FRW universe as
\be\label{eqncphizm0}
\phi(z)=\phi_{0} + \beta (1+z)^{\frac{3}{4}(1+\omega_{\phi})}
\ee
where $\beta=\frac{4}{3f_{0}}{\left[\frac{3H^{2}_{0}(1-\Omega_{m0})}{(1+\omega_{\phi})^{3}}\right]}^{\frac{1}{4}}$.\\ 
With the help of equations (\ref{eqncvzgen}) and (\ref{eqncphizm0}), we have found the form of the potential in terms of $\phi$ as
\be
V(\phi)=V_{0}(\phi -\phi_{0})^4
\ee
where $V_{0}=\frac{3H^2_{0}(1-\Omega_{m 0})(1-3\omega_{\phi})}{4\beta^4}$. Thus, the constant $\omega_{\phi}$ model leads to a quartic potential. For $\phi_{0}=0$, the above potential is similar to the potential used by Linde \cite{gncpotm1} in the context of chaotic inflation. Using the expression for $\phi(z)$, we have also obtained the functional form of $f(\phi)$ as
\be
f(\phi)= {f_{\alpha}{\left(\phi - \phi_{0}\right)}^{-\frac{8}{1 + \omega_{\phi}}}} {{\left(1 + f_{\beta}{\left(\phi - \phi_{0}\right)}^{\frac{4\omega_{\phi}}{1 + \omega_{\phi}}}\right)}^{-2}}
\ee
where, $f_{\alpha}=\frac{f^{4}_{0}{\beta}^{\frac{8}{1 + \omega_{\phi}}}}{H^{4}_{0}\Omega^{2}_{m0}}$ and $f_{\beta}=\frac{(1-\Omega_{m0})}{\Omega_{m0}}\beta^{-\frac{4\omega_{\phi}}{1 + \omega_{\phi}}}$. It is evident from above equation that the function $f(\phi)$ can be arranged as a series expansion in powers of ($\phi - \phi_{0}$).
\subsection{Model-II: Accelerating universe driven by time-dependent EoS parameter for dark energy }\label{tmagnc}\label{gncsecm2}
\begin{figure}
\centering
\includegraphics[width=0.38\textwidth,height=0.22\textheight]{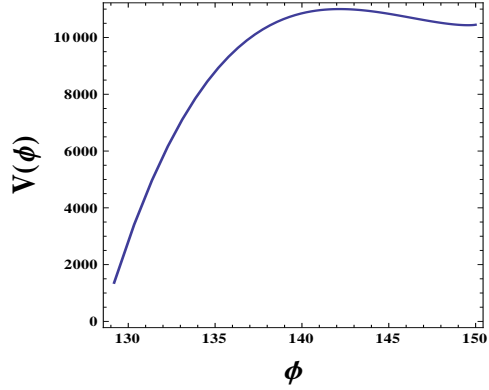}
\caption{\normalsize{\em Plot of $V(\phi)$ vs. $\phi$ for the linear parametrization $ \omega_{\phi}=\omega_{0} + \omega_{1}z$, by assuming $\omega_{0}=-0.95$, $\omega_{1}=0.15$, $\Omega_{m0}=0.3$, $f_{0}=1$, $\phi_{0}=150$ and $H_{0}=72$ km $s^{-1} Mpc^{-1}$.}}
\label{figncvphim1}
\end{figure}

We shall now focus on the second possibility i.e., the EoS parameter $\omega_{\phi}$ is dynamical in nature. For this purpose, in this subsection, we have considered three different choices of $\omega_{\phi}$ to study the model in a more general way.
If $\omega_{\phi}$ is dynamical in nature, then one way to study models going beyond the cosmological constant is by using a particular functional form for the dark energy EoS parameter $\omega_{\phi}(z)$. However, a large number of functional forms for $\omega_{\phi}(z)$ have appeared in the literature \cite{linearwgnc,linearw1gnc,cplnc,cplnc1,logparanc,banc,flinc,ualamnc1p3,ualamnc2p3,
wagncpoly}. Usually, the parametrized form of $\omega_{\phi}(z)$ is written as \cite{revnc0}
\be\label{eqncgenwphiz1}
\omega_{\phi}(z)=\sum_{n=0} \omega_{n}x_{n}(z)
\ee
where $\omega_{n}$'s are arbitrary constants and $x_{n}(z)$'s are functions of redshift $z$.
The numerical values of $\omega_{n}$'s  can be found by fitting it to the observational data. Following first order expansions in equation (\ref{eqncgenwphiz1}), several authors considered many functional forms for $x_{n}(z)$ to investigate the evolution of $\omega_{\phi}(z)$. \\
For example:
\begin{itemize}
\item[i)] $\omega_{\phi}(z)=\omega_{0}=$constant (as we have discussed in model I) for $x_{0}(z)=1$ and $x_{n}=0$ ($n\ge 1$). 
\item[ii)] $\omega_{\phi}(z) = \omega_{0} + \omega_{1} z$ i.e., linear redshift parametrization \cite{linearwgnc,linearw1gnc}, for $x_{n}(z)=z^{n}$ with $n\le 1$.
\item[iii)] $\omega_{\phi}(z) = \omega_{0} + \omega_{1} {\rm log}(1+z)$ i.e., logarithmic parametrization \cite{logparanc}, for $x_{n}(z)=[{\rm log}(1+z)]^{n}$ with $n\le 1$.
\item[iv)] $\omega_{\phi}(z) = \omega_{0} + \omega_{1} \frac{z}{(1+z)}$ i.e., CPL parametrization \cite{cplnc,cplnc1}, for $x_{n}={\left(\frac{z}{1+z}\right)}^{n}$ with $n=1$.
\end{itemize}
and many more. It is worth mentioning that the parametrizations (ii \& iii) diverges at high redshifts, whereas the fourth one blows up in the future, when $z\rightarrow -1$. It should be noted that the assumed parametrization would lead to possible biases in the study of evolution of the DE but in absence of any information regarding the true nature of DE, these parametrizations provide some insight regarding the possible nature of DE component and are worth studying. In this paper, we shall consider two different divergence-free functional forms of $\omega_{\phi}(z)$ which does not diverge in future ($z\rightarrow -1$). In addition, we shall also consider the linear redshift parametrization of $\omega_{\phi}(z)$ for the statistical model comparisons with the divergence-free parametrizations, at low redshifts. In order to explore the evolution of DE, we shall also try to reconstruct the deceleration parameter $q(z)$ using equation (\ref{eqncqz}) for these different choices in section \ref{gncresults}. \\ \\
$\bullet$ {\bf Assumption I}\\ \\
Here, we have considered the linear redshift parametrization of the EoS parameter $\omega_{\phi}$ \cite{linearwgnc,linearw1gnc}, which has the following functional form 
\be\label{eqncp1}
\omega_{\phi}(z) = \omega_{0} + \omega_{1} z
\ee
where, $\omega_{0}$ represents the present value of $\omega_{\phi}$ and the second term measures the variation of $\omega_{\phi}$ with respect to $z$. \\
Inserting for $\omega_{\phi}(z)$ from equation (\ref{eqncp1}) into equation (\ref{eqncrhophigen}), we have obtained
\be\label{eqncrphip1}
\rho_{\phi}(z)=\rho_{\phi 0}(1+z)^{(1+\omega_{0} - \omega_{1})}{\rm exp}(3\omega_{1}z)
\ee
Now equation (\ref{eqnchub}) can be written as
\be\begin{split} &
H^2(z)=H^2_{0}[\Omega_{m0}(1+z)^3 \\ & ~~~~~~~~~~~~~~~~+ (1-\Omega_{m0})(1+z)^{(1+\omega_{0} - \omega_{1})}{\rm exp}(3\omega_{1}z)]
\end{split}
\ee
Now by numerical investigations, we have plotted $V$ as a function of $\phi$ in figure \ref{figncvphim1} by considering $\omega_{0}=-0.95$, $\omega_{1}=0.15$, $\Omega_{m0}=0.3$, $f_{0}=1$ and $\phi_{0}=150$ for this case. Figure \ref{figncvphim1} shows that the potential $V(\phi)$ increases initially but becomes almost flat as $\phi$ increases. The reason behind this seems to be the form of the linear parametrization which is appropriate only for low redshifts ($z<<1$) and diverges for large redshifts. The corresponding expressions for $V(\phi)$ and $f(\phi)$ become approximately equal to  (for details see Appendix B)
\be
V(\phi)\simeq 253.4{\phi}^{3} -75.04 {\phi}^{2} -512.4\phi + 10770 
\ee
and
\be
f(\phi)\simeq 2.53\times 10^{-8} {\rm exp}(32.17\phi)
\ee
\par
\begin{figure}
\centering
\includegraphics[width=0.38\textwidth,height=0.22\textheight]{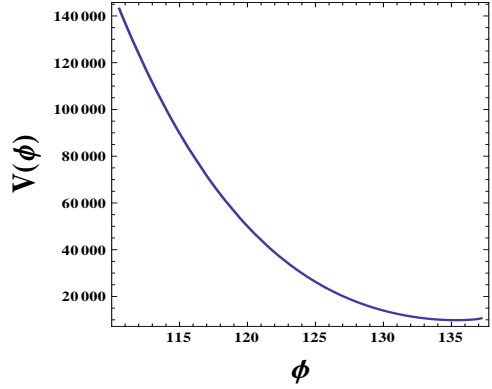}
\caption{\normalsize{\em Plot of $V(\phi)$ vs. $\phi$ for the ansatz given by equation (\ref{eqncp2}). The plot is for $\omega_{2}=-1.25$, $\omega_{3}=5$, $\Omega_{m0}=0.3$, $f_{0}=1$, $\phi_{0}=150$ and $H_{0}=72$ km $s^{-1} Mpc^{-1}$.}}
\label{figncvphim2}
\end{figure}

\begin{figure}
\centering
\includegraphics[width=0.38\textwidth,height=0.22\textheight]{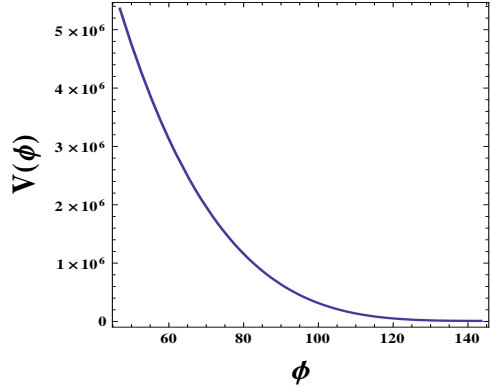}
\caption{\normalsize{\em The variation of $V(\phi)$ with $\phi$ for the ansatz given by equation (\ref{eqncp3}). The plot is for $A_{0}=3.5$, $A_{1}=0.2$, $A_{2}=0.4$, $\Omega_{m0}=0.3$, $f_{0}=1$, $\phi_{0}=150$ and $H_{0}=72$ km $s^{-1} Mpc^{-1}$.}}
\label{figncvphim3}
\end{figure}

 In the context of DE (as it is a late-time phenomenon), the above choice of $\omega_{\phi}(z)$ has been widely used due to its simplicity and we find for the present parametrization of $f$ given in equation (\ref{eqncfans}), the potential comes out to be a polynomial in $\phi$.\\ \\
$\bullet$ {\bf Assumption II}\\ \\
Next, we propose 
\be\label{eqncp2}
\omega_{\phi}(z) = \omega_{2} + \frac{1}{1+ \frac{\omega_{3}}{(1+z)^3}}
\ee
where $\omega_{2}$ and $\omega_{3}$ are arbitrary constants to be fixed by observations. It is easy to see that the EoS parameter reduces to
\bea\label{eqncp2lc}
{\omega_{\phi}(z)} = \left\{\begin{array}{ll} 1+\omega_{2},&$for$\ z\rightarrow +\infty\hspace{1mm} ({\rm early~epoch}),\\\\
\omega_{2} + \frac{1}{1+ \omega_{3}},\ \ \ \ \ \ \ \ \ \ &$for$\
z=0\hspace{1mm} ({\rm present~epoch}), \\\\
\omega_{2},&$for$\ z\rightarrow-1 \hspace{1mm} ({\rm far~future}).
\end{array}\right.
\eea
Thus the above choice of $\omega_{\phi}(z)$ is a bounded function of redshift throughout the entire cosmic evolution and it also overcomes the shortcomings of the linear and CPL parametrizations of $\omega_{\phi}(z)$. Although, this is the main motivation of proposing the ansatz given in equation (\ref{eqncp2}), it can also be thought of as a particular form of equation (\ref{eqncgenwphiz1}) for appropriate choices of $\omega_{n}$'s and $x_{n}(z)$'s.\\ In this case, $\rho_{\phi}(z)$  and $H(z)$ evolve as
\be
\rho_{\phi}(z)=\frac{\rho_{\phi 0}}{1+\omega_{3}}(1+z)^{3(1+\omega_{2})}(\omega_{3} + (1+z)^3)
\ee
where $\rho_{\phi 0}$ is the present value of the scalar field density.
\be\begin{split} &
H^2(z)=H^{2}_{0}[\Omega_{m0}(1+z)^{3}\\ & ~~~~~~~~~~~~~~~+ \frac{(1-\Omega_{m0})}{1+\omega_{3}}(1+z)^{3(1+\omega_{2})}(\omega_{3} + (1+z)^3)]
\end{split}
\ee
For this specific choice, $V(\phi)$ and $f(\phi)$ can be obtained as (see Appendix B)
\be
V(\phi)\simeq 0.11\phi^{4} -58.5\phi^{3} + 12136\phi^{2} -10^{6}\phi + 4\times 10^{7}
\ee
and 
\be
f(\phi)\simeq f_{1}\phi^{6} + f_{2}\phi^{5}+f_{3}\phi^{4} + f_{4}\phi^{3} + f_{5}\phi^{2} + f_{6}\phi + f_{7}
\ee
where $f_{1}=6\times 10^{-15}$, $f_{2}=-5\times 10^{-12}$, $f_{3}=10^{-9}$, $f_{4}=-2\times 10^{-7}$, $f_{5}=2\times 10^{-5}$, $f_{6}=-0.001$ and $f_{7}=0.0205$. These values of $f_{n}$'s have been obtained for $\omega_{2}=-1.25$, $\omega_{3}=5$, $\Omega_{m0}=0.3$, $f_{0}=1$ and $\phi_{0}=150$. In this case, the evolution of the potential $V(\phi)$ is shown in figure \ref{figncvphim2} and we have seen that $V(\phi)$ sharply decreases with $\phi$ from an extremely large value to a fixed value.\\ \\ 
$\bullet$ {\bf Assumption III}\\ \\
The next choice adopted in this paper is suggested by Alam et al. \cite{ualamnc1p3,ualamnc2p3}, which has a functional form
\be\label{eqncp3}
\omega_{\phi}(z)=-1 + \frac{A_{1}(1+z) + 2A_{2}(1+z)^2}{3[A_{0} +A_{1}(1+z) + A_{2}(1+z)^2]}
\ee
This choice is exact and gives the cosmological constant $\omega_{\phi}=-1$ for $A_{1}=A_{2}=0$ and DE models with $\omega_{\phi}=-\frac{1}{3}$ for $A_{0}=A_{1}=0$ and $\omega_{\phi}=-\frac{2}{3}$ for $A_{0}=A_{2}=0$. The above choice mimics a DE model very well and also, it can be viewed as a power law in the redshift dependence of the energy density for DE component. With this choice of $\omega_{\phi}(z)$, equation (\ref{eqncrhophigen}) immediately gives
\be
\rho_{\phi}(z)=\frac{\rho_{\phi 0}}{A_{0} +A_{1} + A_{2}} {\left[A_{0} +A_{1}(1+z) + A_{2}(1+z)^2\right]}
\ee
where $\rho_{\phi 0}$ is the present value of $\rho_{\phi}$.\\
In this case, the Hubble parameter is expressed as
\be\begin{split} &
H^2(z)=H^{2}_{0}[\Omega_{m0}(1+z)^{3}\\ &  ~~~~~~~+ \frac{(1-\Omega_{m0})}{A_{0} +A_{1} + A_{2}} {\left(A_{0} +A_{1}(1+z) + A_{2}(1+z)^2\right)}]
\end{split}
\ee
We have then solved equations (\ref{eqncvzgen}) and (\ref{eqncphiz}) numerically and have plotted $V$ as a function of $\phi$ for some specific values of the model parameters ($A_{0}=3.5$, $A_{1}=0.2$, $A_{2}=0.4$, $\Omega_{m0}=0.3$, $f_{0}=1$ and $\phi_{0}=150$) in figure \ref{figncvphim3}. It is evident from figure \ref{figncvphim3} that the potential $V(\phi)$ always decreases with the scalar field $\phi$. For the present model, $V(\phi)$ and $f(\phi)$ can be explicitly expressed in terms of $\phi$ as (see Appendix B)
\be
V(\phi)\simeq 20920 - 13210\phi - 17300\phi^{2} - 16510\phi^{3}
\ee
and 
\be
f(\phi)\simeq 1.004\times 10^{-53} \phi^{21.23}
\ee
However, in general for Model II (which includes ansatz I, II and III) with the particular choice of equation (\ref{eqncfans}), one can write the potential $V$ as a polynomial in $\phi$ in the following manner
\be\label{gncm2potpoly}
V(\phi) = \sum^{n}_{i=0}V_{i}\phi^{i}
\ee
where, $n>0$, $V_{i}$'s are constants and the values of these parameters are different for different choices of $\omega_{\phi}(z)$. Interestingly, we have found that it is a generalization of other well known potentials (see \cite{revnc0} and the references therein), for example, a constant potential or a power-law potential. We have also found that the parametrization (\ref{eqncfans}) leads to the quantity $f(\phi)$ as exponential, polynomial and power-law in $\phi$ for choices I, II $\&$ III respectively. In the following section, we shall use these choices to discuss the possibility of constraining $\omega_{\phi}(z)$ and $q(z)$ from observations. 
\section{Results}\label{gncresults}
\begin{figure}
\centering
\includegraphics[width=0.38\textwidth,height=0.22\textheight]{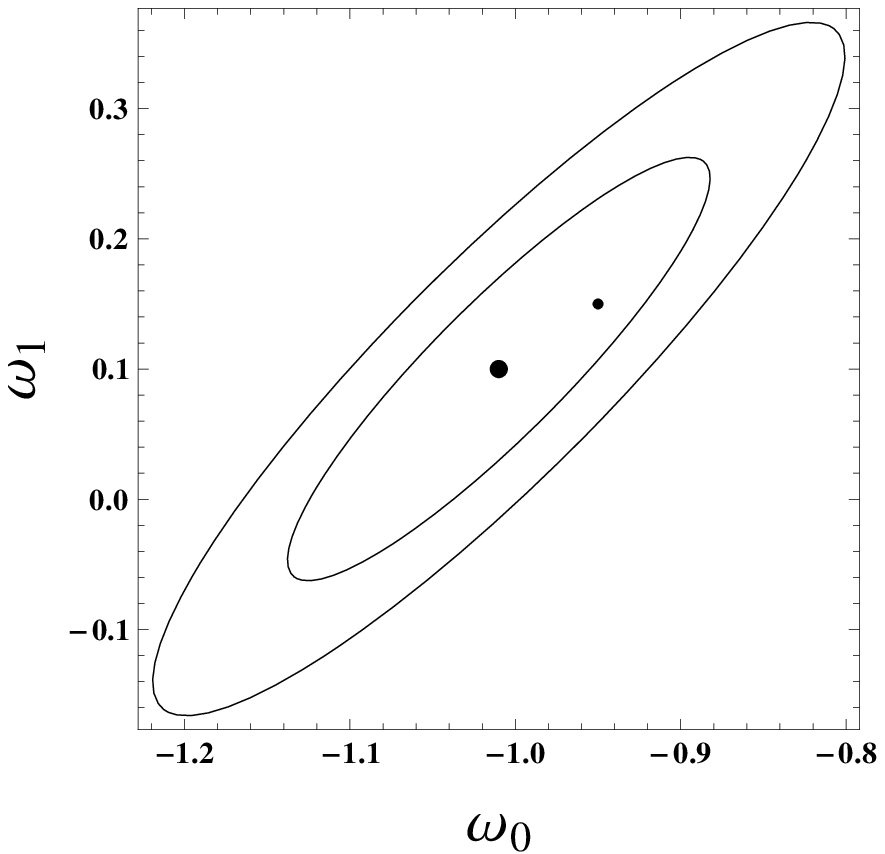}\\
\includegraphics[width=0.38\textwidth,height=0.22\textheight]{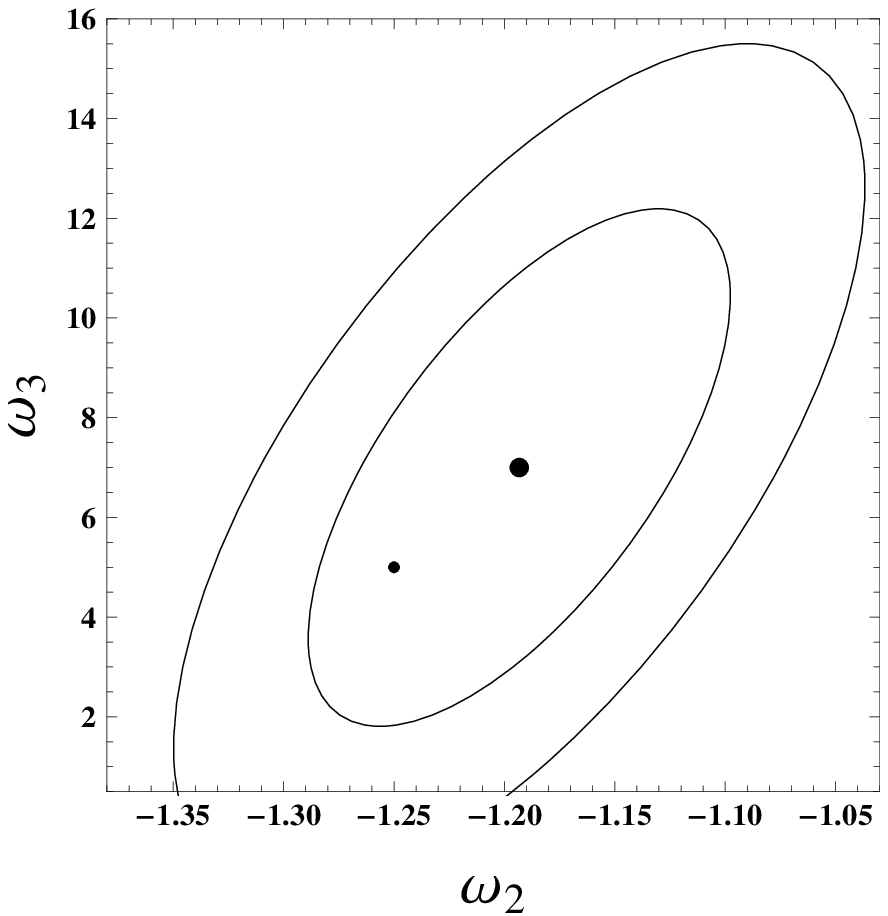}\\
\includegraphics[width=0.38\textwidth,height=0.22\textheight]{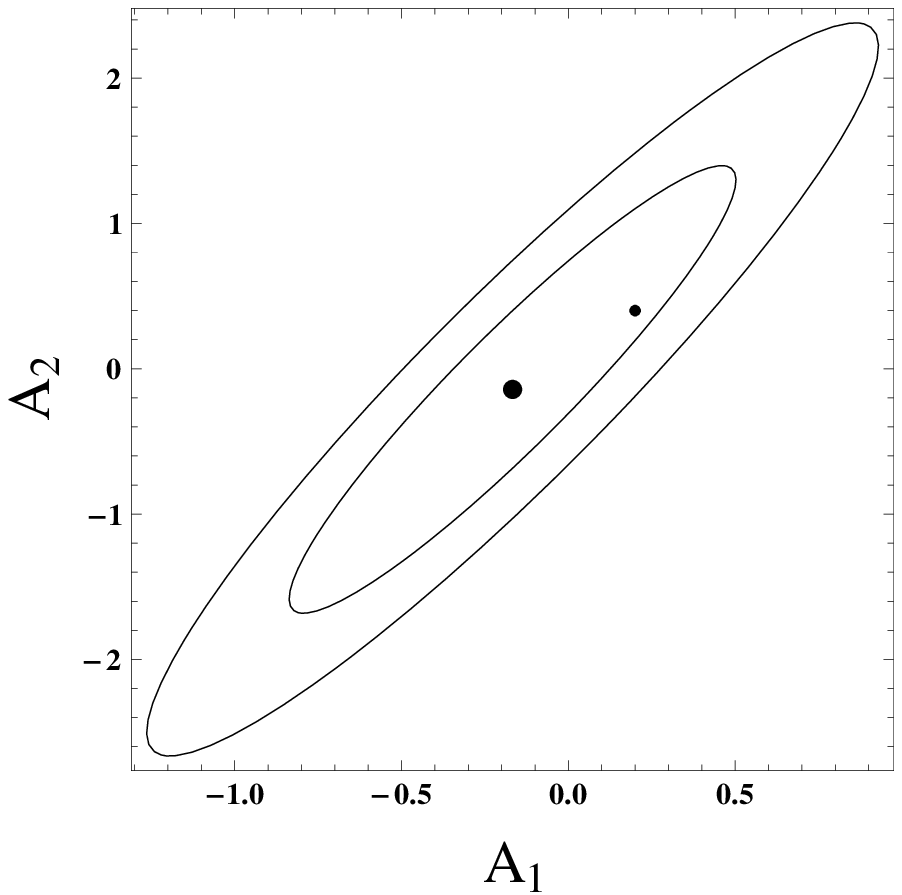}
\caption{\normalsize{\em This figure shows the $1\sigma$ and $2\sigma$ confidence contours for each choice of $\omega_{\phi}(z)$ using the SN Ia $+$ BAO/CMB dataset. The plots are for $\Omega_{m0}=0.3$ (for choices I $\&$ II) and $\Omega_{m0}=0.3$, $A_{0}=3.5$ for choice III. The upper, middle and lower panels represent the $\omega_{0}-\omega_{1}$, $\omega_{2}-\omega_{3}$ and $A_{1}-A_{2}$ parameter space for the choices I, II and III respectively. In each panel, the large dot represents the best-fit values of the model parameters, whereas the small dot represents the chosen values of these parameters in the analytical models (as mentioned in section \ref{tmagnc}). The corresponding $\chi^{2}$ for the best-fit points are displayed in table \ref{tablebfgnc}.}}
\label{figncmodel123c}
\end{figure}

\begin{figure}
\centering
\includegraphics[width=0.38\textwidth,height=0.22\textheight]{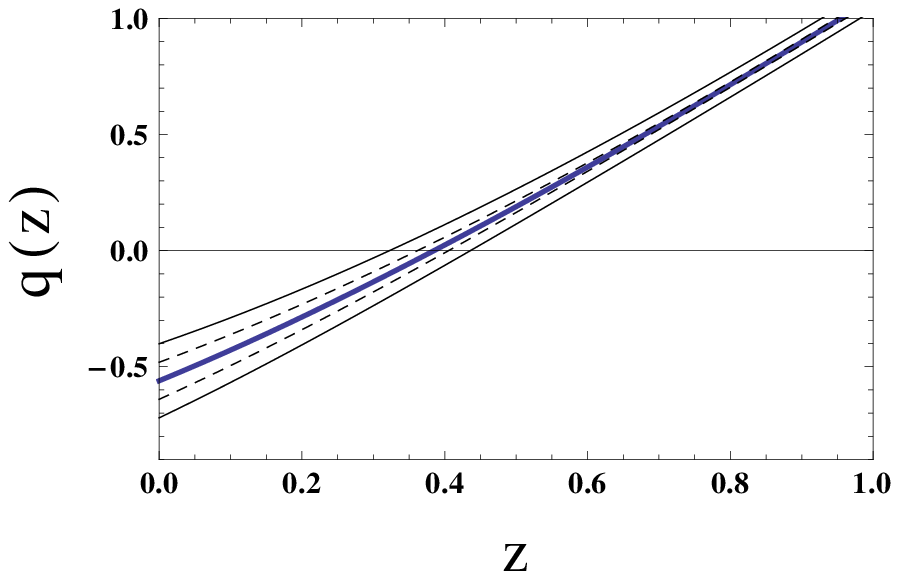}\\
\includegraphics[width=0.38\textwidth,height=0.22\textheight]{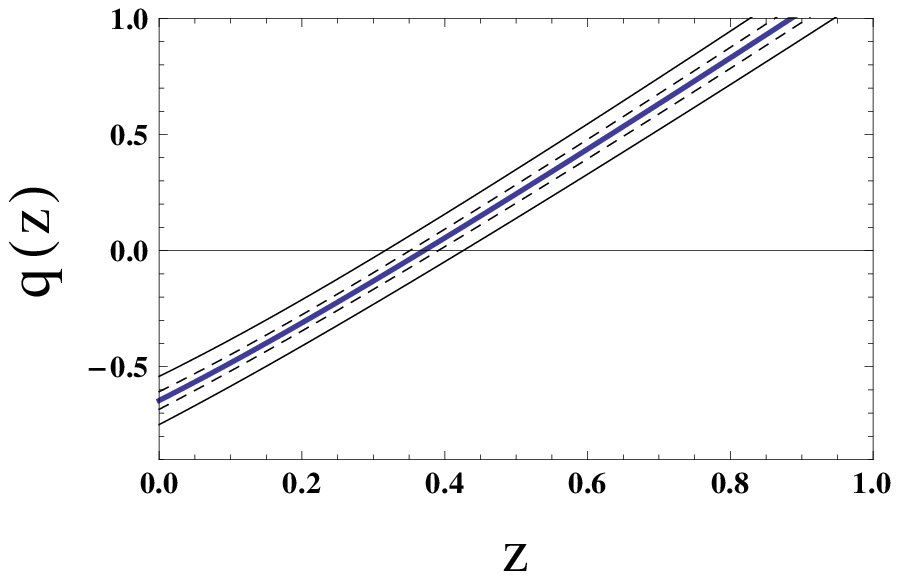}\\
\includegraphics[width=0.38\textwidth,height=0.22\textheight]{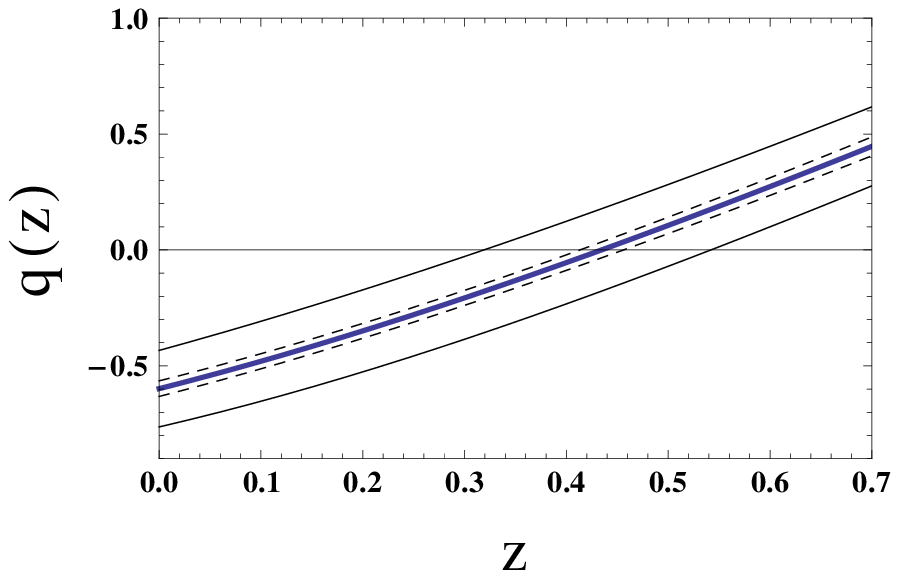}
\caption{\normalsize{\em This figure shows the evolution of $q(z)$ with redshift $z$ for the choices I (upper panel), II (middle panel) and III (lower panel) respectively. The reconstruction is done using the SN Ia $+$ BAO/CMB dataset by assuming $\Omega_{m0}=0.3$ for choices I, II and $\Omega_{m0}=0.3$, $A_{0}=3.5$ for choice III. In each panel, the thick solid line shows the best-fit curve, the dashed lines represent the $1\sigma$ confidence level, and the thin lines represent the $2\sigma$ confidence
level around the best-fit. Also, in each panel, the horizontal line indicates $q(z)=0$.}}
\label{figncmodel123qz}
\end{figure}

Following the statistical analysis (see Appendix A), in this section, we have presented the fitting results for different choices of the EoS parameter for DE. Figure \ref{figncmodel123c} shows the $1\sigma$ and $2\sigma$ confidence contours for each choice (I, II and III) using the SN Ia $+$ BAO/CMB dataset. 
The best fit values of the model parameters and $\omega_{\phi}(z=0)$ for these different choices are given in table \ref{tablebfgnc}. 
\begin{table*}
\caption{\it Best fit values for various model parameters for the analysis of SN Ia $+$ BAO/CMB dataset. Here, $\omega_{\phi}(z=0)$ represents the present value of the EoS parameter $\omega_{\phi}(z)$ in the best-fit models. For this analysis, we have considered $\Omega_{m0}=0.3$ (for choices I $\&$ II) and $\Omega_{m0}=0.3$, $A_{0}=3.5$ for choice III.}
\begin{center}
\begin{tabular}{|c|c|c|c|c|c|}
\hline
Choice   & \multicolumn{2}{|c|}{Best fit values of} &$\omega_{\phi}(z=0)$&${\chi^{2}_{m}}$\\
&\multicolumn{2}{|c|}{model parameters}&&\\
\hline
I&$\omega_{0}=-1.01$&$\omega_{1}=0.10$&$-1.01$&$599.90$\\
\hline
II&$\omega_{2}=-1.19$&$\omega_{3}=7$&$-1.06$&$565.43$\\
\hline
III&$A_{1}=-0.16$&$A_{2}=-0.14$&$-1.04$&$564.86$\\
\hline
\end{tabular}
\label{tablebfgnc}
\end{center}
\end{table*}
Using those best-fit values, we have reconstructed the deceleration parameter $q(z)$ for each model and the results are plotted in figure \ref{figncmodel123qz}. 
It is evident from figure \ref{figncmodel123qz} that $q(z)$ shows a smooth transition from a decelerated ($q>0$) to an accelerated ($q<0$) phase of expansion of the universe at the transition redshift $z_{t}=0.38$ (for ansatz I), $0.36$ (for ansatz II) and $0.43$ (for ansatz III) for the best-fit models. These results are in good agreement with those obtained by several authors from various other considerations \cite{gncztagr1,gncztagr2,gncztcl}.\\
\par
\begin{figure}
\centering
\includegraphics[width=0.38\textwidth,height=0.22\textheight]{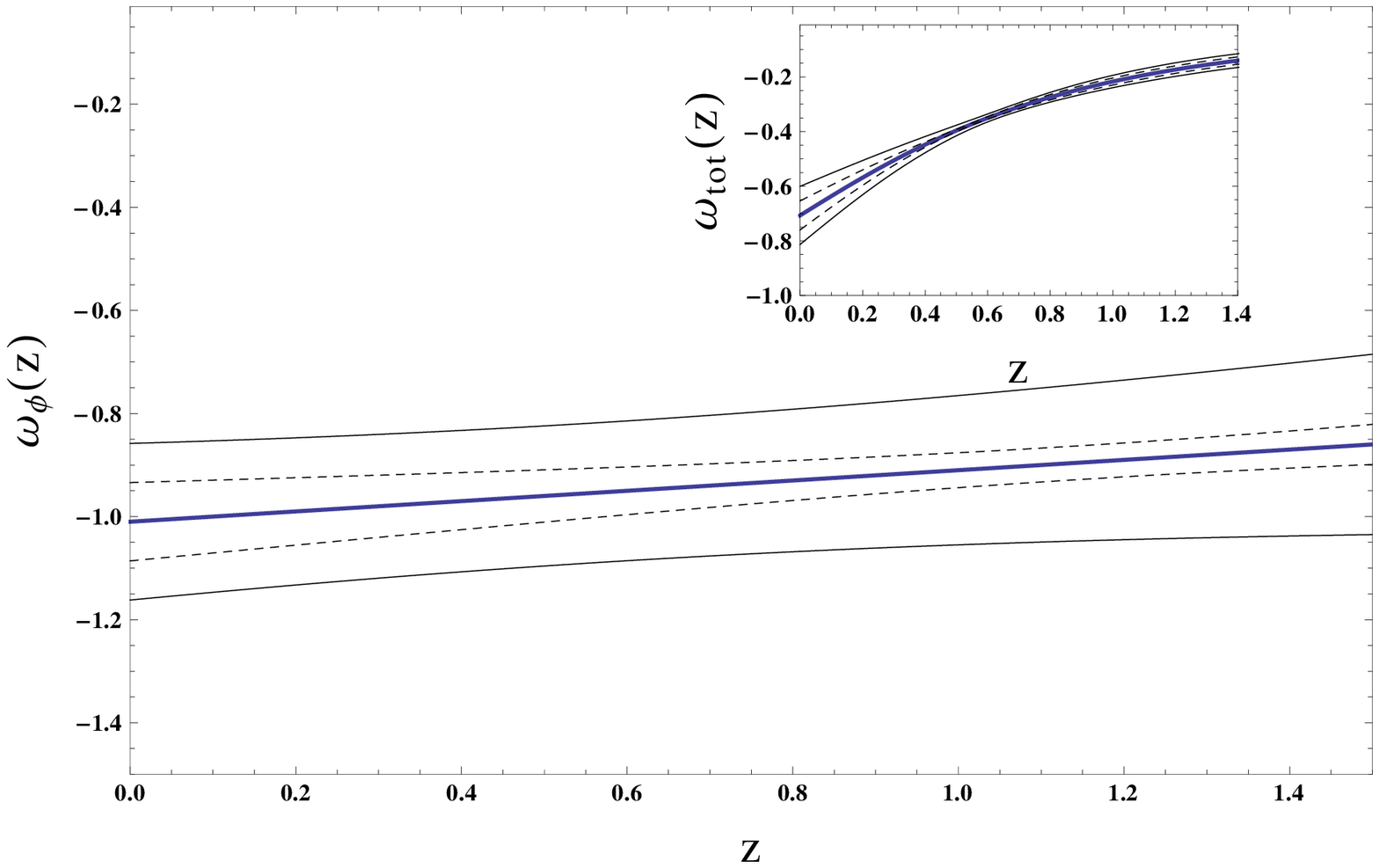}\\
\includegraphics[width=0.38\textwidth,height=0.22\textheight]{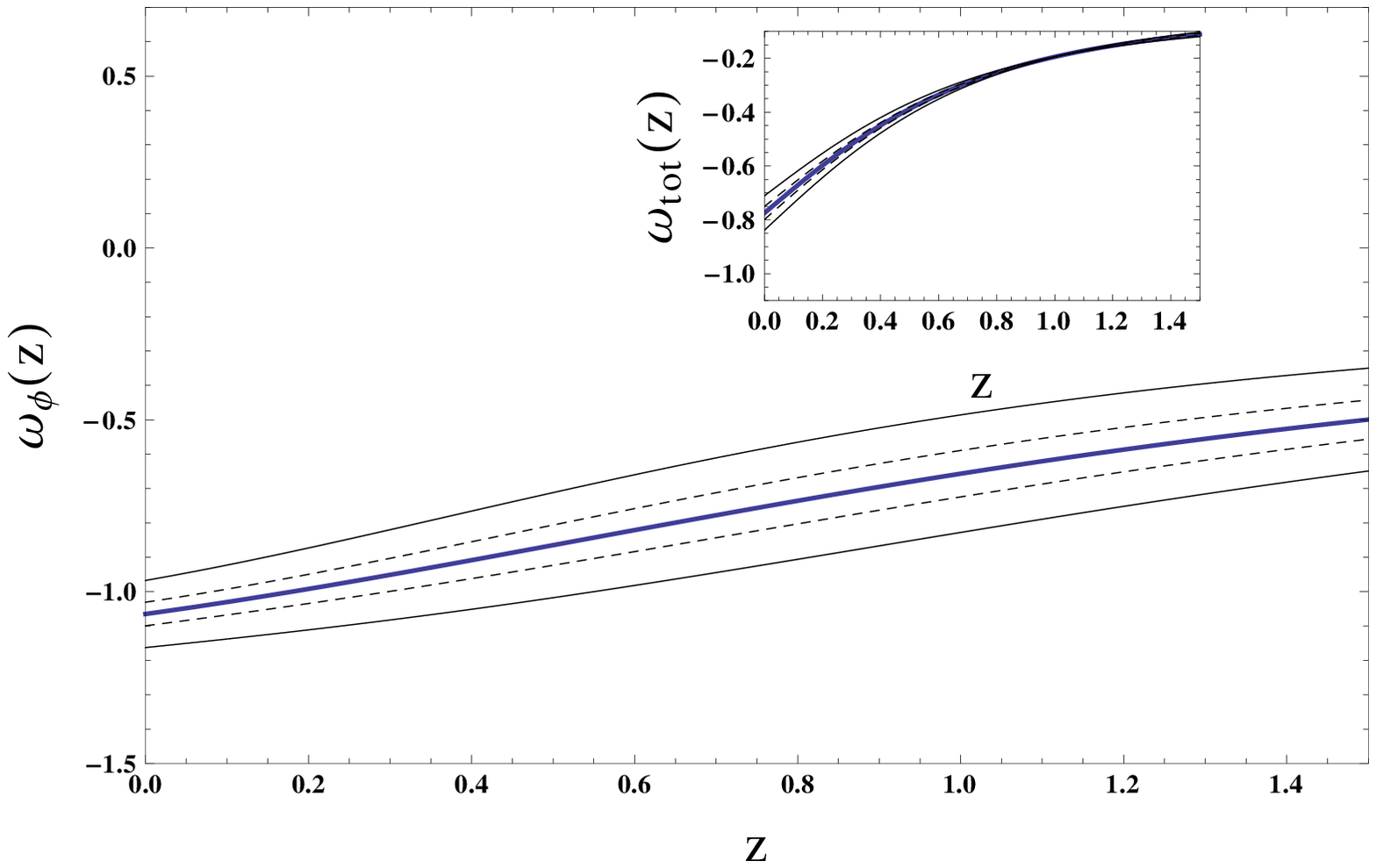}\\
\includegraphics[width=0.38\textwidth,height=0.22\textheight]{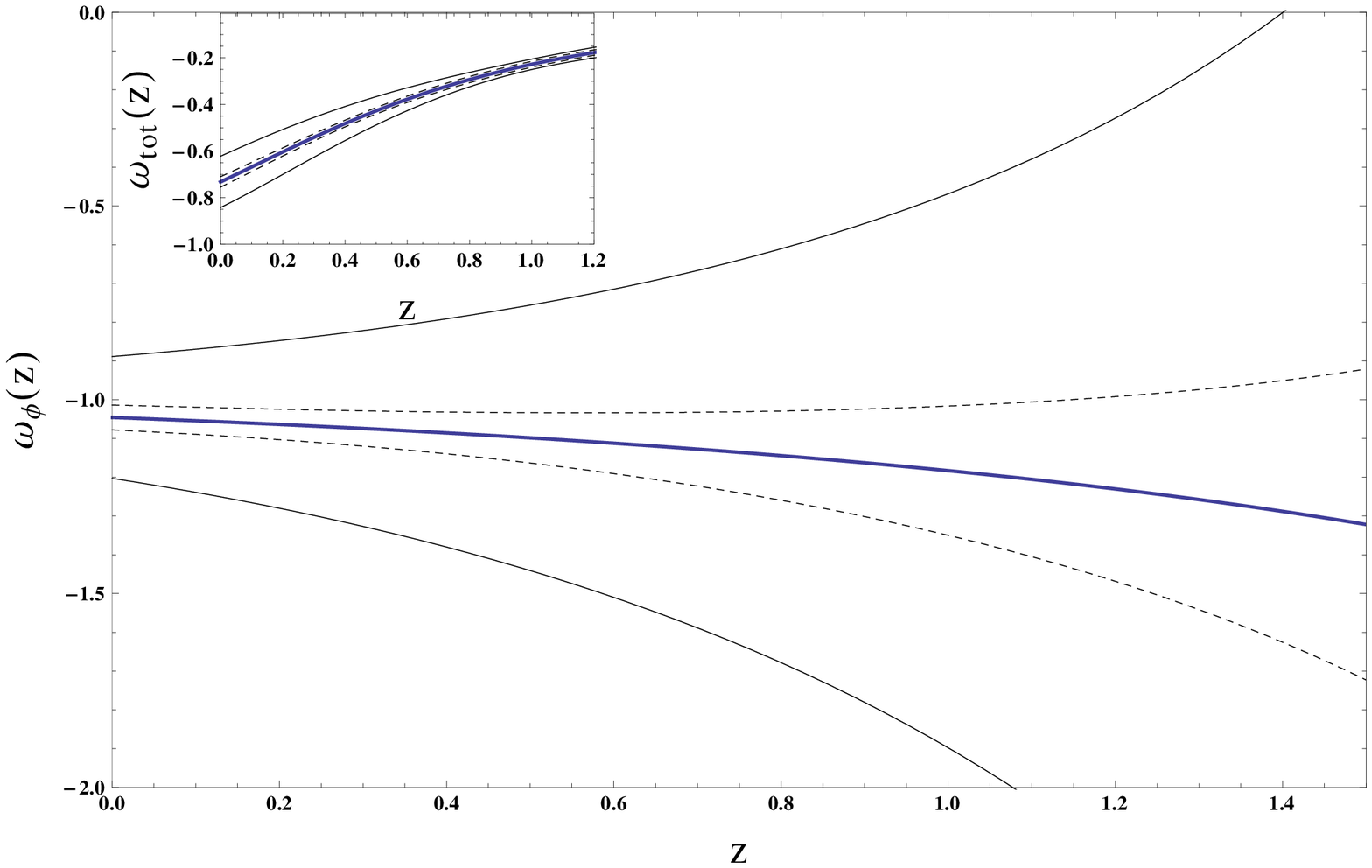}\\
\caption{\normalsize{\em The upper and middle panel represents the plot of $\omega_{\phi}(z)$ vs. $z$ with $\Omega_{m0}=0.3$ for ansatzs I and II respectively. The lower panel corresponds to the evolution of $\omega_{\phi}(z)$ for the ansatz III. This plot is for $\Omega_{m0}=0.3$ and $A_{0}=3.5$. Also, in each panel, the inset diagram shows the evolution of the total EoS parameter $\omega_{tot}(z)$ with $z$ for these ansatzs. The thick solid line shows the best-fit curve, the dashed lines represent the $1\sigma$ confidence level, and the thin lines represent the $2\sigma$ confidence level around the best-fit.}}
\label{figncmodel123wphiz}
\end{figure}

 Furthermore, we have also shown the reconstructed evolution history of the EoS parameter in figure \ref{figncmodel123wphiz} for each choice of $\omega_{\phi}(z)$.
We have also plotted the total EoS parameter, which is defined as $\omega_{tot}(z)=\frac{p_{\phi}}{\rho_{\phi} + \rho_{m}}$ , as a function of $z$ for these choices (see inset diagram of figure \ref{figncmodel123wphiz}). From table \ref{tablebfgnc}, we have found that the current values of $\omega_{\phi}(z)$ for the best-fit DE models are very close to $-1$, i.e., the models do not deviate very far from the $\Lambda$CDM model ($\omega_{\Lambda}=-1$) at the present epoch. However, as indicated  in table \ref{tablebfgnc}, the present parametrized model favours a phantom model ($\omega_{\phi}<-1$) in $2\sigma$ limit and thus requires further attention.
\section{Conclusions}\label{gncconclusions}
In this work, we have studied various non-canonical scalar field DE models in a spatially flat, homogeneous and isotropic FRW space-time. In this framework, we have obtained the general solutions of the field equations for different choices of the EoS parameter. For completeness, we have also investigated how the joint analysis of SN Ia $+$ BAO/CMB dataset constrains the redshift evolutions of $q(z)$ and $\omega_{\phi}(z)$ for different choices of $\omega_{\phi}(z)$ (as given in Model II).  In figure \ref{figncmodel123c}, we have also shown the $1\sigma$ and $2\sigma$ contour plots of the pairs ($\omega_{0} , \omega_{1}$) (upper panel), ($\omega_{2} , \omega_{3}$) (middle panel) and ($A_{1} , A_{2}$)  (lower panel) for the ansatzs I, II and III respectively. In this analysis, we have also calculated the best-fit values of the free parameters (as shown by large dot in figure \ref{figncmodel123c}) and it has been found that the chosen values of these parameters (which were chosen for solving the parametric relations in Appendix B) are well fitted within the $1\sigma$ confidence contour (as shown by small dot in figure \ref{figncmodel123c}).  \\
\par We have shown that the deceleration parameter $q$ undergoes a smooth transition from its deceleration phase ($q>0$, at high $z$) to an acceleration phase ($q<0$, at low $z$) for all of the considered parametrized models. However, as mentioned in the previous section, the value of $z_{t}$, where the signature flip of $q$ (from the decelerating to an accelerating expansion phase) takes place has been calculated and the results obtained are consistent with the present day cosmological observations. From the SN Ia $+$ BAO/CMB analysis, we have also found $q(z=0)=-0.56$, $-0.64$ and $-0.60$ for ansatzs I, II and III respectively which also agree very well with the recent observational results.\\ 
\par From table \ref{tablebfgnc}, we have observed that the EoS parameter $\omega_{\phi}(z=0)\approx -1$, but slightly less than $-1$ for all three choices (as discussed in section \ref{gncresults}). As we have seen $\omega_{\phi}(z=0)\approx -1$, hence our models do not deviate very far from the $\Lambda$CDM model (see also figure \ref{figncmodel123wphiz}), which is currently known as the standard model for modern cosmology. In order to gain more physical insight into these time evolutions of the EoS parameter, we have also plotted the reconstructed total EoS parameter $\omega_{tot}(z)$ in figure \ref{figncmodel123wphiz} (see inset diagram of figure \ref{figncmodel123wphiz}). For each choice, this figure shows that $\omega_{tot}(z)$ attains the require value of $-\frac{1}{3}$ around $z=0.62$ (within $1\sigma$ confidence level) and remains always greater than $-1$ upto the present epoch. These scenarios also agree very well with the observational data.\\
\par However, the models presented here are restricted because the form of $f(\phi)$ chosen was ad-hoc (as given in equation (\ref{eqncfans})) and did not follow from any principle. In this regard, we have mentioned earlier that we make this choice in order to close the system of equations. With this choice of $f(\phi)$, we have derived the form of the potential $V(\phi)$ in terms of $\phi$ for different models. We have found that Model I leads to a quartic potential, whereas Model II leads to  a polynomial potential for each choice of $\omega_{\phi}(z)$. We have seen that, with a suitable choice of $V_{i}$'s for the potential (as given in equation (\ref{gncm2potpoly})), it is possible to reproduce the other well known potentials in the context of DE. However, many possibilities are opened up to accommodate a physically viable potential for other parametrization of $f(\phi)$ or $f(H)$. Finally, we would like to emphasize that all the considered models provide a deceleration for high redshift and an acceleration for low redshift as required for the structure formation of the universe. However, these results are completely independent of any choice of $f(\phi)$. With the increase of more good quality observational data at the low, intermediate and high redshifts, the constraints on $z_{t}$ (or $q(z)$) and $\omega_{\phi}(z)$ are expected to get improved in the near future. 
\section{Acknowledgements}One of the authors (AAM) is thankful to Govt. of India for financial support through Maulana Azad National Fellowship. SD wishes to thank IUCAA, Pune for associateship program.\\ \\
{\bf Appendix A: Data analysis method}\\ \\
In this section, we shall fit the theoretical models with the recent observational datasets from the type Ia supernova (SN Ia), the baryonic acoustic oscillations (BAO) and the cosmic microwave background (CMB) data surveying. For completeness, we shall briefly summarize each of the datasets.\\ \\
$\bullet$ {\bf SN Ia dataset:}\\ \\
In this paper, we have considered recently released Union2.1 compilation \cite{sn1agnc}, which totally contains 580 data points with redshift ranging from 0.015 to 1.414. To constraint cosmological parameter using SN Ia dataset, the $\chi^2$ function is defined as (see ref. \cite{sndatamethodgnc})
\be\label{chisquare} 
\chi^2_{SN} = A_{SN} - \frac{B^2_{SN}}{C_{SN}}
\ee
where $A_{SN}$, $B_{SN}$ and $C_{SN}$ are defined as follows
\bea
A_{SN} = \sum^{580}_{i=1} \frac{[{\mu}^{obs}(z_{i}) - {\mu}^{th}(z_{i})]^2}{\sigma^2_{i}},\\
B_{SN} = \sum^{580}_{i=1} \frac{[{\mu}^{obs}(z_i) - {\mu}^{th}(z_{i})]}{\sigma^2_{i}},
\eea
and
\be 
C_{SN} = \sum^{580}_{i=1} \frac{1}{\sigma^2_{i}}
\ee 
where $\mu^{obs}$ represents the observed distance modulus while $\mu^{th}$ the theoretical one and $\sigma_{i}$ is the error associated with each data point.\\ \\
$\bullet$ {\bf BAO/CMB dataset:}\\ \\
Next, we have used BAO \cite{bao1gnc,bao2gnc,bao3gnc} and CMB \cite{cmbgnc} measurements data to obtain the BAO/CMB constraints on the model parameters. For the BAO/CMB dataset, the details of methodology for obtaining the constraints on model parameters are described in ref. \cite{goistrignc}. The $\chi^2$ function for this dataset is defined as 
\be
\chi^2_{BAO/CMB} = X^{T}C^{-1}X 
\ee
where the transformation matrix ($X$) and the inverse covariance matrix ($C^{-1}$) are given in ref. \cite{goistrignc}.\\
Finally, the total $\chi^{2}$ for these observational datasets is given by
\be
\chi^{2}_{total} = \chi^{2}_{SN} +\chi^{2}_{BAO/CMB}
\ee
For this analysis, we have used the normalized Hubble parameter which is defined as $h(z)=\frac{H(z)}{H_{0}}$. The quantity $h(z)$ contains only three free parameters, namely, $\Omega_{m0}$, $\omega_{i}$ and $\omega_{j}$ for assumption I ( $i=0$, $j=1$) and II ($i=2$, $j=3$). For the sake of simplicity, we have reduced the three dimensional parameter space ($\Omega_{m0}$, $\omega_{i}$, $\omega_{j}$) into the two dimensional plane ($\omega_{i}$, $\omega_{j}$) by fixing $\Omega_{m0}$ to some constant value. On the other hand, $h(z)$ contains four free parameters ($\Omega_{m0}$, $A_{0}$, $A_{1}$ and $A_{2}$) for assumption III. In this case, we have also reduced the four dimensional parameter space ($\Omega_{m0}$, $A_{0}$, $A_{1}$, $A_{2}$) into the two dimensional plane ($A_{1}$, $A_{2}$) by fixing $\Omega_{m0}$ and $A_{0}$ to some constant values. Now, we can deal with only two free parameters for each ansatz and will perform $\chi^{2}$ analysis of the SN Ia $+$ BAO/CMB dataset. The values of the model parameters at which  $\chi^{2}_{m}$ (the minimum value of $\chi^{2}$ function) is obtained are the best-fit values of these parameters for the joint analysis of the observational datasets from SN Ia, BAO and CMB measurements.\\ \\
{\bf Appendix B: Solutions of $\phi(z)$, $V(z)$ and $f(z)$ for each choice of $\omega_{\phi}(z)$ (I, II and III)}\\ \\
In this section, we shall briefly extend our discussion regarding the solutions of $\phi(z)$, $V(z)$ and $f(z)$ for different choices of $\omega_{\phi}(z)$ used in Model II.\\ \\
$\bullet$ {\bf Assumption I:}\\ \\
For this choice, we have obtained the evolution of $\phi(z)$ by integrating equation (\ref{eqncphiz}) numerically and is given by
\be\begin{split} &\label{eqncappphizp1}
\phi(z)=\phi_{0}\\ & + \alpha_{1}\frac{G_{1}(z){F_{1}} {\left[\frac{5}{4},\alpha_{3},-\frac{1}{4}, \frac{9}{4},\frac{1+\omega_{0}+\omega_{1}z}{\omega_{1}(1+z)},\frac{\alpha_{4}(1+\omega_{0}+\omega_{1}z)}{(1+z)}\right]}}{(1+z)^{\frac{3\omega_{1}-3\omega_{0}-4}{4}}}
\end{split}
\ee
where $G_{1}(z)=(1+\omega_{0} + \omega_{1}z)^{\frac{5}{4}}{\left(-\frac{1+\omega_{0}-\omega_{1}}{\omega_{1}(1+z)}\right)}^{\frac{1}{4}(1+3\omega_{0}-3\omega_{1})}$, $\alpha_{1}=\frac{4 (1+\omega_{0}-\omega_{1}) {\left(\frac{3H^{2}_{0}(1-\Omega_{m0})}{\alpha_{2}}\right)}^{\frac{1}{4}} }{5f_{0}\omega^{2}_{1}}$, $\alpha_{2}=\frac{(1+\omega_{0}-\omega_{1})}{\omega_{1}(2+3\omega_{0})}$, $\alpha_{3}=\frac{3}{4}(3+\omega_{0}-\omega_{1})$ and $\alpha_{4}=\frac{3\omega_{1}-1}{\omega_{1}(2+3\omega_{0})}$. It is worth mentioning that we have considered ${\rm exp}(\omega_{1}z)\approx 1+\omega_{1}z$ in equation (\ref{eqncrphip1}) to compute the integration numerically, otherwise it becomes very difficult to get a solution for $\phi(z)$.\\ 
Now, using equation (\ref{eqncvzgen}), we have found the potential (in terms of $z$) as
\be\label{eqncappvzp1}
V(z)=V_{01}(1-3\omega_{0}-3\omega_{1}z)(1+z)^{3(1+\omega_{0}-\omega_{1})}{\rm exp}(\omega_{1}z)
\ee
where $V_{01}=\frac{3H^{2}_{0}(1-\Omega_{m0})}{4}$. From equation (\ref{eqncfans}), we have also obtained
\be\label{eqncappfzp1}
f(z)=\frac{ {(f_{0}/H_{0})}^{4}}{{\left[\Omega_{m0}(1+z)^3 + (1-\Omega_{m0})(1+z)^{(1+\omega_{0} - \omega_{1})}{\rm exp}(3\omega_{1}z)\right]}^{2}}
\ee
$\bullet$ {\bf Assumption II:}\\ \\
Similarly, for ansatz II, the functional forms of $\phi(z)$, $V(z)$ and $f(z)$ can be expressed as
\be\begin{split} &\label{eqncappphizp2}
\phi(z)=\phi_{0}+ \beta_{1} (1+z)^{\frac{3}{4}(2+\omega_{2})} \\ &
\times {{}_{2}F_{1}}{\left[-\frac{1}{4},-\frac{(2+\omega_{2})}{4},\frac{(2-\omega_{2})}{4},-\frac{\beta_{2}}{(1+z)^3}\right]},
\end{split}
\ee
\be\begin{split} &\label{eqncappvzp2}
V(z)=V_{02} (1+z)^{3(1+\omega_{2})} {\left(\omega_{3} + (1+z)^{3}\right)}\\ &
\times{\left[1 - 3\omega_{2} -\frac{3(1+z)^{3}}{\omega_{3} + (1+z)^{3}}\right]},
\end{split}
\ee
and 
\be\label{eqncappfzp2}
f(z)=\frac{ {(f_{0}/H_{0})}^{4}}{{\left[\Omega_{m0}(1+z)^{3} + \frac{(1-\Omega_{m0})}{1+\omega_{3}}(1+z)^{3(1+\omega_{\phi})}\right]}^{2}}
\ee
where $\beta_{1}=-\frac{4}{f_{0}} {\left(\frac{H^{2}_{0}(1-\Omega_{m0})}{27(1+\omega_{3})(2+\omega_{2})^{3}}\right)}^{\frac{1}{4}}$, $\beta_{2}=\frac{(1+\omega_{2})\omega_{3}}{(2+\omega_{2})}$ and $V_{02}=\frac{3H^{2}_{0}(1-\Omega_{m0})}{4(1+\omega_{3})}$.\\ \\
$\bullet$ {\bf Assumption III:}\\ \\
For ansatz III, we have also obtained
\be\begin{split} &\label{eqncappphizp3}
\phi(z)=\phi_{0}\\ & + \frac{ G_{2}(z) {\left(-8A_{2} -\frac{4A_{1}}{1+z} + G_{3}(z){{}_{2}F_{1}}{\left[\frac{1}{2},\frac{3}{4},\frac{3}{2},-\frac{A_{1}}{2A_{2}(1+z)}\right]}\right)}}{f_{0} {\left(4A_{2} + \frac{2A_{1}}{1+z}\right)} },
\end{split}
\ee
\be\begin{split} &\label{eqncappvzp3}
V(z)=V_{03} (A_{0} + A_{1}(1+z) + A_{2}(1+z)^{2})\\ &
~~~~~~~~~~~~~~\times {\left[4 - \frac{A_{1}(1+z) + 2A_{2}(1+z)^{2}}{A_{0} + A_{1}(1+z) + A_{2}(1+z)^{2}}\right]}
\end{split}
\ee
and
\be\label{eqncappfzp3}
f(z)=\frac{ {(f_{0}/H_{0})}^{4}}{[\Omega_{m0}(1+z)^{3} + \frac{(1-\Omega_{m0})}{A_{0} +A_{1} + A_{2}} {\left(A_{0} +A_{1}(1+z) + A_{2}(1+z)^2\right)}]^{2}}
\ee
where $G_{2}(z)={\left[\frac{H^{2}_{0}(1-\Omega_{m0})(1+z)^{2}{\left(2A_{2} + \frac{A_{1}}{1+z}\right)}}{A_{0} + A_{1} + A_{2}}\right]}^{\frac{1}{4}}$,\\ $G_{3}(z)=\frac{2^{\frac{1}{4}}A_{1} {\left(2+\frac{A_{1}}{A_{2}(1+z)}\right)}^{\frac{3}{4}}}{1+z}$ and $V_{03}=\frac{3H^{2}_{0}(1-\Omega_{m0})}{4(A_{0} + A_{1} + A_{2})}$.\\
\par To reconstruct $V(\phi)$, we have proceeded as follows. One can easily find that it is not possible to express $V(\phi)$ in terms of $\phi$ explicitly, because $\phi(z)$ poses very complicated form for each choice of $\omega_{\phi}(z)$. These equations only give a parametric representation of $V(\phi)$, which cannot be solved analytically. Therefore, one can plot the potential $V(\phi)$ against $\phi$ for some arbitrary values of the model parameters. After this, one can obtain the form of $V(\phi)$ by using a fitting function to fit the corresponding plot. Following this procedure, we have plotted $V(\phi)$ as a function of $\phi$ for these choices (I, II and III),  which are shown in figure \ref{figncvphim1}, \ref{figncvphim2} and \ref{figncvphim3}.  Similarly, using the parametric relations $[f(z),$ $\phi(z)]$ for each choice of $\omega_{\phi}(z)$, we have also obtained the form of $f(\phi)$ by numerical method for some given values of the model parameters.\\

\end{document}